\documentclass[12pt,letterpaper]{article}
\usepackage{graphicx}
\usepackage{accents}

\def\title#1{{\bf\Large #1}}
\def\sec#1{\vspace{1.00\baselineskip} \noindent {\bf\large #1}
\vspace{0.25\baselineskip}}

\def\al{\alpha}
\def\bt{\beta}
\def\ga{\gamma}

\def\de{\delta}

\def\De{\Delta}

\def\la{\lambda}
\def\La{\Lambda}

\def\pd{\partial}
\def\ph{\phi}
\def\ps{\psi}

\def\si{\sigma}

\def\th{\theta}

\def\ze{\zeta}

\def\d{\dagger}
\def\<{\langle}
\def\>{\rangle}

\def\tsum{{\textstyle\sum}}

\def\d{\dagger}
\def\tr{{\rm tr}}
\def\diag{{\rm diag}}

\def\ba{\begin{eqnarray}}
\def\ea{\end{eqnarray}}
\def\be{\begin{equation}}
\def\ee{\end{equation}}

\begin{document}


\begin{center}

\vspace*{1.0\baselineskip}

\title{Introduction to the symplectic group Sp(2)}

\vspace{1.0\baselineskip}

C. J. McKinstrie \\

{\it\small Independent Photonics Consultant, Manalapan, NJ 07726, United States}

\vspace{0.50\baselineskip}

M. V. Kozlov

{\it\small Center for Preparatory Studies, Nazarbayev University, Astana 010000, Kazakhstan}

\vspace{0.50\baselineskip}

Abstract \\

\vspace{0.50\baselineskip}

\parbox[]{6.5in}{\small  In this article, we derive and discuss the properties of the symplectic group Sp(2), which arises in Hamiltonian dynamics and ray optics. We show that a symplectic matrix can be written as the product of a symmetric dilation matrix and a rotation matrix, in either order. A symplectic matrix can be written as the exponential of a generating matrix, and there is a one-to-one relation between the coefficients of the symplectic and generating matrices. We also discuss the adjoint and Schmidt decompositions of a symplectic matrix, and the product of two symplectic matrices. The results of this article have applications in many subfields of physics.}

\end{center}

\newpage

\sec{1. Introduction and basic concepts}

Group theory has many applications in the physical sciences, because it facilitates the modeling, understanding and classification of physical phenomena  \cite{jos82,tin03}.
In this article, we review the properties of the symplectic group Sp(2), which is the set of $2 \times 2$ real matrices with determinant 1. The concepts and techniques required to study this simple group can also be used to determine the properties of more complicated matrix groups, which we will discuss in a future article.

Symplectic transformations originate in Hamiltonian dynamics \cite{gol01,tay05}. Let $q$ and $p$ be the (normalized) displacement and momentum, respectively, of a simple harmonic oscillator. Then the oscillator dynamics are governed by the Hamiltonian
\be H = (p^2 + q^2)/2, \label{1.1} \ee
together with the Hamilton equations
\be d_t q = \pd H/\pd p, \ \ d_t p = -\pd H/\pd q, \label{1.2} \ee
where $d_t$ is a time derivative. By combining Eqs. (\ref{1.1}) and (\ref{1.2}), one obtains the dynamical equations
\be d_t q = p, \ \ d_t p = -q. \label{1.3} \ee

Let $X = [q, p]^t = [x_1, x_2]^t$ be a variable (coordinate) vector. Then Eqs. (\ref{1.3}) can be written in the matrix form
\be d_t X = JX, \label{1.4} \ee
where the coefficient (structure) matrix
\be J = \left[\begin{array}{cc} 0 & 1 \\ -1 & 0 \end{array}\right]. \label{1.5} \ee
Notice that $J^t = -J$, $J^2 = -I$ and $J^tJ = I$, so $J$ is orthogonal.

The inner product $X^tX =$ $x_1^2 + x_2^2$ is the (normalized) energy of the oscillator. Equation (\ref{1.4}) conserves energy, because
\be d_t X^tX = X^tJ^tX + X^tJX = X^t(J^t + J)X = 0. \label{1.6} \ee
Now let $Y = [y_1, y_2]^t$ be another coordinate vector. Then the bilinear (quadratic) form $X^tJY = x_1y_2 - x_2y_1$ is the cross-product of $X$ and $Y$, whose magnitude is the area in state space of the parallelogram with sides $X$ and $Y$. Equation (\ref{1.4}) also conserves area, because
\be d_t X^tJY = X^tJ^tJY + X^tJ^2Y = X^t(I - I)Y = 0. \label{1.7} \ee

The solution of Eq. (\ref{1.4}) can be written in the input--output (IO) form
\be X(t) = T(t)X(0), \label{1.8} \ee
where $T(t) = \exp(Jt)$ is the transfer (Green) matrix and $J$ is its generator \cite{cra83}. By using the identity $J^2 = -I$ to exponentiate $Jt$ [Eq. (\ref{3.23})], one finds that
\be T(t) = \left[\begin{array}{cc} \cos t & \sin t \\ -\sin t& \cos t \end{array}\right]. \label{1.9} \ee
Notice that $T^t = T^{-1}$, so $T$ is an orthogonal (rotation) matrix. In the transfer-matrix formalism, the inner product $X^tY$ is conserved, because
\be X^t(t)Y(t) = X^tT^tTY = X^t Y. \label{1.10} \ee
The cross-product $X^tJY$ is also conserved, because
\be X^t(t)JY(t) = X^tT^tJTY = X^tT^tTJY = X^tJY. \label{1.11} \ee
A similar matrix formalism applies to systems with more complicated Hamiltonians \cite{mck13} or more degrees of freedom \cite{cra83}. The conservation of area (volume) in state space is an important property of Hamiltonian systems.

In dynamics, transfer matrices arise as the exponentials of generators. In ray optics, they arise directly. The distance of a ray from the optical axis and the slope of the ray (multiplied by the index of refraction) play the roles of displacement and momentum, respectively \cite{buc93,bor99}. By using the laws of geometric optics (reflection and refraction), one can determine the transfer matrices for simple elements, such as lenses and spaces \cite{ger94,pea15}. These matrices are also symplectic. The transfer matrix for a composite system is the product of the transfer matrices for its constituent elements. In ray optics, area (volume) in state space is called \'etendue (extent).

The key part of Eq. (\ref{1.11}) is the symplectic equation $T^tJT = J$. In the oscillator example, $T$ satisfies this equation because it commutes with $J$. However, it is not the most general matrix that satisfies the equation.
In Sec. 2, we study the properties of symplectic matrices, which form a group under multiplication \cite{taw95}. We show that every symplectic matrix $M$ can be written as the product of a dilation matrix (which stretches in one direction and squeezes in the orthogonal direction) and a rotation matrix. It is useful to know that every symplectic transformation has a simple physical interpretation. The dilation matrix is symmetric and is specified by two real parameters, the dilation magnitude and the orientation angle. The rotation matrix is asymmetric and is specified by one real parameter, the rotation angle.

In the same way that $T(t) = \exp(Jt)$, every symplectic matrix can be written in the form $M(t) = \exp(Gt)$, where $G$ is a generating matrix and $t$ is a parameter. In Sec. 3, we study the properties of generating matrices, which form a vector space under addition \cite{taw91}. We show that every generating matrix can be written as the combination $G_1k_1 + G_2k_2 + G_3k_3$, where $G_1$ -- $G_3$ are fundamental (basis) generators and $k_1$ -- $k_3$ are real generator coefficients. We derive formulas for the matrix elements in terms of the generator coefficients, and the generator coefficients in terms of the matrix elements.

The product of two symplectic matrices is also a symplectic matrix. In Sec. 4, we consider the product of two dilation matrices. We relate the dilation and angle parameters of the product matrix to the dilation and angle parameters of the constituent matrices. The composition of two dilations is not necessarily another dilation.

In Sec. 5, we discuss ray-tracing matrices briefly, to illustrate the occurrence and use of symplectic matrices, generators and decompositions. We state the transfer matrices for propagation and focusing, and describe how to produce ideal magnification. Every ray transformation can be decomposed into propagation, magnification and focusing.

In Sec. 6, we summarize the main results of this article. In Apps. A and B, we discuss the adjoint and Schmidt decompositions of symplectic matrices. These decompositions are generalizations of the spectral decomposition, which only applies to symmetric matrices. In App. C, we discuss the product of two arbitrary symplectic matrices.
The results of this article have applications in classical dynamics \cite{gol01,tay05} and ray optics \cite{ger94,pea15}, and analogous results have applications in quantum optics \cite{wod85,ger01} and special relativity \cite{jac99,tip08}.

\newpage

\sec{2. Symplectic matrices}

Let $M$ be a real $2 \times 2$ matrix. Then $M$ is symplectic if it satisfies the equation
\be M^tJM = J, \label{2.1} \ee
where $J$ was defined in Eq. (\ref{1.5}) \cite{kim83,sat86,gil08}. Trivial examples of such matrices include the identity matrix $I$ and $J$ itself (because $J^tJ = I$). If Eq. (\ref{2.1}) is satisfied, then $J^t(M^tJM) = J^tJ = I$. Hence, $M$ has the inverse
\be M^{-1} = J^tM^tJ. \label{2.2} \ee
Conversely, if Eq. (\ref{2.2}) is satisfied, then $J = JM^{-1}M = M^tJM$. Hence, Eqs. (\ref{2.1}) and (\ref{2.2}) are equivalent. When one tries to establish the properties of symplectic matrices, one can use whichever equation is more convenient.

For example, if $M$ is symplectic, then $(M^{-1})^{-1} = J^{-1}(M^t)^{-1}(J^t)^{-1} = J^t(M^{-1})^tJ$. In addition, $J = MM^{-1}J = MJ^tM^tJ^2 = MJM^t$. Hence, $M^{-1}$ and $M^t$ are also symplectic. The latter result allows one to write Eq. (\ref{2.1}) in the alternative form $MJM^t = J$. If $M_1$ and $M_2$ are symplectic, then $(M_2M_1)^tJ(M_2M_1) = M_1^t(M_2^tJM_2)M_1 = M_1^tJM_1 = J$. Hence, $M_2M_1$ is also symplectic. By using the principle of induction, one can extend this result to the product of an arbitrary number of symplectic matrices. It follows from the laws of matrix multiplication that $M_3(M_2M_1) = (M_3M_2)M_1 = M_3M_2M_1$.

To summarize these properties, the set of symplectic matrices is closed under multiplication, which is associative. The set contains the identity matrix, and every member of the set has an inverse. Hence, the set of symplectic matrices is a group under multiplication \cite{taw95}.

Let $\al$, $\bt$, $\ga$ and $\de$ be the elements (components) of $M$. Then
\ba M^tJM &=&\left[\begin{array}{cc} \al & \ga \\ \bt & \de \end{array}\right]
\left[\begin{array}{cc} 0 & 1 \\ -1 & 0 \end{array}\right]
\left[\begin{array}{cc} \al & \bt \\ \ga & \de \end{array}\right] \nonumber \\
&= &\left[\begin{array}{cc} \al & \ga \\ \bt & \de \end{array}\right]
\left[\begin{array}{cc} \ga & \de \\ -\al & -\bt \end{array}\right] \nonumber \\
&= &\left[\begin{array}{cc} 0 & \al\de - \bt\ga \\ \bt\ga - \al\de & 0 \end{array}\right]. \label{2.3} \ea
Hence, $M$ is symplectic if and only if $\det(M) = 1$. (This condition is not sufficient in higher dimensions.) $M$ has four real components, which are subject to one real constraint equation, so the symplectic group Sp(2,R) is a three-parameter group. The special linear group SL(2,R) is the set of real $2 \times 2$ matrices that have determinant 1. Hence, Sp(2,R) equals SL(2,R). In higher dimensions, the symplectic group is a subgroup of the special linear group. (Henceforth, we will omit R.) It follows from Eq. (\ref{2.2}) that the inverse matrix
\ba M^{-1} &=&\left[\begin{array}{cc} 0 & -1 \\ 1 & 0 \end{array}\right]
\left[\begin{array}{cc} \al & \ga \\ \bt & \de \end{array}\right]
\left[\begin{array}{cc} 0 & 1 \\ -1 & 0 \end{array}\right] \nonumber \\
&= &\left[\begin{array}{cc} 0 & -1 \\ 1 & 0 \end{array}\right]
\left[\begin{array}{cc} -\ga & \al \\ -\de & \bt \end{array}\right] \nonumber \\
&= &\left[\begin{array}{cc} \de & -\bt \\ -\ga & \al \end{array}\right]. \label{2.4} \ea
Formula (\ref{2.4}) is the standard formula for the inverse.

Three nontrivial examples of symplectic matrices are
\be \left[\begin{array}{cc} e^\la & 0 \\ 0 & e^{-\la} \end{array}\right], \ \
\left[\begin{array}{cc} \cosh\ze & \sinh\ze \\ \sinh\ze & \cosh\ze \end{array}\right], \ \
\left[\begin{array}{cc} \cos\th & -\sin\th \\ \sin\th & \cos\th \end{array}\right]. \label{2.5} \ee
In Eq. (\ref{2.5}), the first matrix $D$ represents dilation (stretching and squeezing), the second matrix $B$ represents Lorentz boosting and the third matrix $R$ represents active rotation.
It is easy to verify that the product of two such dilation, boost or rotation matrices is also a dilation, boost or rotation matrix, with parameter $\la_1 + \la_2$, $\ze_1 + \ze_2$ or $\th_1 + \th_2$, respectively.

To make a dilation in the direction $(c, s)$, where $c = \cos\th$ and $s = \sin\th$, one rotates the coordinate system, dilates (stretches) in the new $x$-direction and rotates the coordinate system back to its original orientation \cite{kim83}. Hence,
\ba M(\la, \th) &= &\left[\begin{array}{cc} c & -s \\ s & c \end{array}\right]
\left[\begin{array}{cc} e^\la & 0 \\ 0 & e^{-\la} \end{array}\right]
\left[\begin{array}{cc} c & s \\ -s & c \end{array}\right] \nonumber \\
&= &\left[\begin{array}{cc} c & -s \\ s & c \end{array}\right]
\left[\begin{array}{cc} ce^\la & se^\la \\ -se^{-\la} & ce^{-\la} \end{array}\right] \nonumber \\
&= &\left[\begin{array}{cc} c^2e^\la + s^2e^{-\la} & cs(e^\la-e^{-\la}) \\ cs(e^\la-e^{-\la}) & s^2e^\la + c^2e^{-\la} \end{array}\right] \nonumber \\
&= &\left[\begin{array}{cc} \cosh(\la) + \sinh(\la)\cos(2\th) & \sinh(\la)\sin(2\th) \\ \sinh(\la)\sin(2\th) & \cosh(\la) - \sinh(\la)\cos(2\th) \end{array}\right]. \label{2.6} \ea
Notice that the dilation matrix is symmetric and has determinant 1 (as it should do). Notice also that
\be M(\la,\pi/4) = \left[\begin{array}{cc} \cosh\la & \sinh\la \\ \sinh\la & \cosh\la \end{array}\right]. \label{2.7} \ee
Hence, $B$ is not independent of $D$ and $R$.

One can rewrite Eq. (\ref{2.6}) in the compact notation
\be M(\la, \th) = R_a(\th)D(\la)R_a^{-1}(\th) = R_p^{-1}(\th)D(\la)R_p(\th), \label{2.8} \ee
where $a$ and $p$ denote active and passive rotations, respectively. $M$ is a similarity transform of $D$ and has the same eigenvalues as $D$. Furthermore,
\ba R(\th_2)M(\la,\th_1)R^{-1}(\th_2) &= &R(\th_2)R(\th_1)D(\la)R^{-1}(\th_1)R^{-1}(\th_2) \nonumber \\
&= &R(\th_2 + \th_1)D(\la)R^{-1}(\th_2 + \th_1). \label{2.9} \ea
A similarity transform of $M$ is a dilation of the same magnitude, in a different direction.

The dilation matrix in Eq. (\ref{2.6}) involves only 2 free parameters ($\la$ and $\th$), so it is not the most general symplectic matrix. If one were to pre- or post-multiply a dilation matrix by a rotation matrix, then the product matrix would involve three parameters ($\la$, $\th_1$ and $\th_2$). In fact, this product is the general form of a symplectic matrix.

Let $M$ be an arbitrary symplectic matrix and $R$ be a rotation matrix. Then the product
\ba R^{-1}M &= &\left[\begin{array}{cc} c & s \\ -s & c \end{array}\right]
\left[\begin{array}{cc} \al & \bt \\ \ga & \de \end{array}\right] \nonumber \\
&= &\left[\begin{array}{cc} c\al + s\ga & c\bt + s\de \\ -s\al + c\ga & -s\bt + c\de \end{array}\right]. \label{2.11} \ea
If $c\bt + s\de = -s\al + c\ga$, then the product matrix $N$ is symmetric. One can make it so by setting
\be t = s/c = (\ga - \bt)/(\al + \de), \label{2.12} \ee
in which case
\be N = \left[\begin{array}{cc} \al' & \bt' \\ \bt' & \de' \end{array}\right]  = \left[\begin{array}{cc} \al'' + \de'' & \bt' \\ \bt' & \al'' - \de'' \end{array}\right], \label{2.13} \ee
where $\al'' = (\al' + \de')/2$ and $\de'' = (\al' - \de')/2$.
[One can derive formulas for $\al'$, $\bt'$ and $\de'$ in terms of $\al$, $\bt$, $\ga$ and $\de$ by rewriting the cosine and sine in Eq. (\ref{2.11}) in terms of the tangent in Eq. (\ref{2.12}).]
The determinant condition $(\al'')^2 - (\de'')^2 - (\bt')^2 = 1$ allows one to write $\al'' = C$, $\de'' = Sc$ and $\bt' = Ss$, where $C^2 - S^2 = 1$ and $c^2 + s^2 = 1$. Hence, $M = RN$, where
\be N = \left[\begin{array}{cc} C + Sc & Ss \\ Ss & C - Sc \end{array}\right]. \label{2.14} \ee
The symmetric matrix in Eq. (\ref{2.14}) has the same form as the dilation matrix in Eq. (\ref{2.6}).
It follows from this result that $M = RN = RNR^{-1}R = N'R$, so the rotation matrix that appears in the decomposition can follow or precede the dilation matrix.
The decomposition theorem can be stated in terms of active or passive rotations ($R_a = R_p^{-1}$).

As we stated above, the product of any number of symplectic matrices is symplectic, so the rotation matrices in Eq. (\ref{2.6}) need not be equal. Let $M = R_2DR_1^{-1}$, where $D = D(\la)$ and $R_i = R(\th_i)$. Then, written explicitly,
\ba M &= &\left[\begin{array}{cc} c_2 & -s_2 \\ s_2 & c_2 \end{array}\right]
\left[\begin{array}{cc} C + S & 0 \\ 0 & C - S \end{array}\right]
\left[\begin{array}{cc} c_1 & s_1 \\ -s_1 & c_1 \end{array}\right] \nonumber \\
&= &\left[\begin{array}{cc} c_2 & -s_2 \\ s_2 & c_2 \end{array}\right]
\left[\begin{array}{cc} (C + S)c_1 & (C + S)s_1 \\ -(C - S)s_1 & (C - S)c_1 \end{array}\right] \nonumber \\
&= &\left[\begin{array}{cc} C(c_2c_1 + s_2s_1) + S(c_2c_1 - s_2s_1) & S(s_2c_1 + c_2s_1) - C(s_2c_1 - c_2s_1) \\
S(s_2c_1 + c_2s_1) + C(s_2c_1 - c_2s_1) & C(c_2c_1 + s_2s_1) - S(c_2c_1 - s_2s_1) \end{array}\right] \nonumber \\
&= &\left[\begin{array}{cc} Cc_- + Sc_+ & Ss_+ - Cs_- \\ Ss_+ + Cs_- & Cc_- - Sc_+ \end{array}\right], \label{2.31} \ea
where $C = \cosh(\la)$, $S = \sinh(\la)$, $c_\pm = \cos(\th_\pm)$, $s_\pm = \sin(\th_\pm)$ and $\th_\pm = \th_2 \pm \th_1$. It is easy to verify that $\det(M) = 1$.

One can write $M$ in the alternative forms
\be \left[\begin{array}{cc} \al & \bt \\ \ga & \de \end{array}\right]
= \left[\begin{array}{cc} \al' + \de' & \bt' - \ga' \\ \bt' + \ga' & \al' - \de' \end{array}\right], \label{2.32} \ee
where the primed coefficients
\be \al' = (\al + \de)/2, \ \ \bt' = (\bt + \ga)/2, \ \ \ga' = (\ga - \bt)/2, \ \ \de' = (\al - \de)/2. \label{2.33} \ee
By comparing Eqs. (\ref{2.31}) and (\ref{2.32}), one finds that
\be Cc_- = \al', \ \ Ss_+ = \bt', \ \ Cs_- = \ga', \ \ Sc_+ = \de'. \label{2.34} \ee
It follows from Eqs. (\ref{2.33}) and (\ref{2.34}) that
\ba C^2 &= &(\al')^2 + (\ga')^2 \ = \ (\al^2 + \bt^2 + \ga^2 + \de^2 + 2)/4, \label{2.35} \\
S^2 &= &(\bt')^2 + (\de')^2 \ = \ (\al^2 + \bt^2 + \ga^2 + \de^2 - 2)/4. \label{2.36} \ea
Notice that $C^2 - S^2 = 1$ (as it should do). With $C$ and $S$ known, the dilation parameter
\be \la = \log(C + S). \label{2.37} \ee
The angles are specified implicitly by the equations
\ba \tan(\th_+) &= &\bt'/\de' \ = \ (\bt + \ga)/(\al - \de), \label{2.38} \\
\tan(\th_-) &= &\ga'/\al' \ = \ (\ga - \bt)/(\al + \de). \label{2.39} \ea
Notice that $\th_-$ is the rotation angle required to symmetrize the matrix [Eq. (\ref{2.12})].
The sum and difference of these angles are $\th_+ + \th_- = 2\th_2$ and $\th_+ - \th_- = 2\th_1$, respectively. By using the trigonometric formulas
\be \tan(\th_+ \pm \th_-) = {t_+ \pm t_- \over 1 \mp t_+t_-}, \label{2.40} \ee
where $t_\pm = \tan(\th_\pm)$, one finds that
\ba \tan(2\th_1) &= &{\al'\bt' - \ga'\de' \over \al'\de' + \bt'\ga'} \ = \ {2(\al\bt + \ga\de) \over \al^2 - \bt^2 + \ga^2 - \de^2}, \label{2.41} \\
\tan(2\th_2) &= &{\al'\bt' + \ga'\de' \over \al'\de' - \bt'\ga'} \ = \ {2(\al\ga + \bt\de) \over \al^2 + \bt^2 - \ga^2 - \de^2}. \label{2.42} \ea

Formula (\ref{2.31}) specifies the matrix components $\al$, $\bt$, $\ga$ and $\de$ in terms of the dilation and angle parameters $\la$, $\th_1$ and $\th_2$, whereas formulas (\ref{2.37}), (\ref{2.41}) and (\ref{2.42}) specify the parameters in terms of the components. The latter formulas are valid for components with arbitrary values. Hence, every symplectic matrix can be written as the product of two rotation matrices and a diagonal dilation matrix. Furthermore, $R_2DR_1^{-1} = (R_2R_1^{-1})(R_1DR_1^{-1}) = R_{21}N_1$, where $R_{21} = R(\th_2 - \th_1)$ is a rotation matrix and $N_1$ is a symmetric dilation matrix. Alternatively, $R_2DR_1^{-1} = (R_2DR_2^{-1})(R_2R_1^{-1}) = N_2R_{21}$. The preceding analysis is a constructive proof of the decomposition theorem.

Although we will not study higher-dimensional symplectic groups in detail, some results for Sp(2$n$) are worth mentioning.
First, let $M$ be a $2n \times 2n$ symplectic matrix. Then
\be M = \left[\begin{array}{cc} A & B \\ C & D \end{array}\right], \label{2.51} \ee
where $A$, $B$, $C$ and $D$ are $n \times n$ matrices. The symplectic equation requires that 
\ba M^tJM &= &\left[\begin{array}{cc} A^t & C^t \\ B^t & D^t \end{array}\right]
\left[\begin{array}{cc} 0 & I \\ -I & 0 \end{array}\right]
\left[\begin{array}{cc} A & B \\ C & D \end{array}\right] \nonumber \\
&= &\left[\begin{array}{cc} A^t & C^t \\ B^t & D^t \end{array}\right]
\left[\begin{array}{cc} C & D \\ -A & -B \end{array}\right] \nonumber \\
&= &\left[\begin{array}{cc} A^tC - C^tA & A^tD - C^tB \\ B^tC - D^tA & B^tD - D^tB \end{array}\right] \label{2.52} \ea
equals $J$. Hence, $A^tC$ and $B^tD$ are symmetric, and
\be A^tD - C^tB = I, \ \ D^tA - B^tC = I. \label{2.53} \ee
Furthermore, the inverse matrix
\ba M^{-1} &= &\left[\begin{array}{cc} 0 & -I \\ I & 0 \end{array}\right]
\left[\begin{array}{cc} A^t & C^t \\ B^t & D^t \end{array}\right]
\left[\begin{array}{cc} 0 & I \\ -I & 0 \end{array}\right] \nonumber \\
&= &\left[\begin{array}{cc} 0 & -I \\ I & 0 \end{array}\right]
\left[\begin{array}{cc} -C^t & A^t \\ -D^t & B^t \end{array}\right] \nonumber \\
&= &\left[\begin{array}{cc} D^t & -B^t \\ -C^t & A^t \end{array}\right]. \label{2.54} \ea
This result generalizes the inverse formula for a $2 \times 2$ matrix. It does not apply to a general $2n \times 2n$ matrix, so Sp(2$n$) does not equal, but is a subgroup of, SL(2$n$).

Second, a real $2n \times 2n$ matrix has $4n^2$ real parameters. The reqirements that $A^tC$ and $B^tD$ are symmetric each provide $n(n - 1)/2$ constraints. The first of Eqs. (\ref{2.53}) provides $n^2$ constraints. The second equation is the transpose of the first, so it provides no new constraints. Hence, the number of free parameters is $4n^2 - n(n - 1) - n^2 = n(2n +1)$. This number increases as the square of $n$.

Third, according to the Schmidt decomposition theorem \cite{hor13}, every real matrix has the decomposition $M = QDP^t$, where $D$ is diagonal and non-negative, and $P$ and $Q$ are orthogonal. [Equation (\ref{2.31}) is such a decomposition.] Hence, $M = (QP^t)(PDP^t) = RN_p$, where $R$ is orthogonal and $N_p$ is symmetric. Alternatively, $M = (QDQ^t)(QP^t) = N_qR$.
This result generalizes the result for Sp(2), which was derived above from first principles. The Schmidt decomposition of a symplectic matrix is discussed in detail in App. B.

\newpage

\sec{3. Generating matrices}

Consider the differential equation
\be d_t X = GX, \label{3.1} \ee
where $t$ is a time-like variable and $G$ is a coefficient matrix. Equation (\ref{3.1}) is a generalization of the Hamilton equation (\ref{1.4}). Its solution can be written in the IO form
\be X(t) = T(t)X(0), \label{3.2} \ee
where $T(t) = \exp(Gt)$ is the transfer (Green) matrix \cite{cra83}. Suppose that $Y$ satisfies the same equation. As we showed in Sec. 1, Hamiltonian dynamics conserve the cross-product $X^tJY$. Under what conditions does Eq. (\ref{3.1}) also conserve the cross-product? It follows from Eq. (\ref{3.1}) that
\ba d_t (X^tJY) &= &(X^tG^t)JY + X^tJ(GY) \nonumber \\
&= &X^t(G^tJ + JG)Y. \label{3.3} \ea
Hence, $X^tJY$ is conserved if and only if
\be G^tJ + JG = 0. \label{3.4} \ee
By writing Eq. (\ref{3.4}) in terms of components, one finds that $\tr(G) = 0$. Hence, $\det(T) = e^{\tr(G)t} = 1$ (as it should do).

The $2 \times 2$ matrices
\be G_1 = \left[\begin{array}{cc} 1 & 0 \\ 0 & -1 \end{array}\right], \ \ 
G_2 = \left[\begin{array}{cc} 0 & 1 \\ 1 & 0 \end{array}\right], \ \ 
G_3 = \left[\begin{array}{cc} 0 & -1 \\ 1 & 0 \end{array}\right] \ \ \label{3.11} \ee
all have zero trace (and $G_3 = -J$). They are called generators (of the transfer matrix). Notice that $G_1^2 = G_2^2 = I$ and $G_3^2 = -I$.
It is easy to verify that
\ba G_1G_2 &= &\left[\begin{array}{cc} 1 & 0 \\ 0 & -1 \end{array}\right]
\left[\begin{array}{cc} 0 & 1 \\ 1 & 0 \end{array}\right] \ = \ \left[\begin{array}{cc} 0 & 1 \\ -1 & 0 \end{array}\right]
\ = \ -G_3, \label{3.12} \\
G_2G_1 &= &\left[\begin{array}{cc} 0 & 1 \\ 1 & 0 \end{array}\right]
\left[\begin{array}{cc} 1 & 0 \\ 0 & -1 \end{array}\right] \ = \ \left[\begin{array}{cc} 0 & -1 \\ 1 & 0 \end{array}\right]
\ = \ G_3, \\
G_2G_3 &= &\left[\begin{array}{cc} 0 & 1 \\ 1 & 0 \end{array}\right]
\left[\begin{array}{cc} 0 & -1 \\ 1 & 0 \end{array}\right] \ = \ \left[\begin{array}{cc} 1 & 0 \\ 0 & -1 \end{array}\right]
\ = \ G_1, \\
G_3G_2 &= &\left[\begin{array}{cc} 0 & -1 \\ 1 & 0 \end{array}\right]
\left[\begin{array}{cc} 0 & 1 \\ 1 & 0 \end{array}\right] \ = \ \left[\begin{array}{cc} -1 & 0 \\ 0 & 1 \end{array}\right]
\ = \ -G_1, \\
G_3G_1 &= &\left[\begin{array}{cc} 0 & -1 \\ 1 & 0 \end{array}\right]
\left[\begin{array}{cc} 1 & 0 \\ 0 & -1 \end{array}\right] \ = \ \left[\begin{array}{cc} 0 & 1 \\ 1 & 0 \end{array}\right]
\ = \ G_2, \\
G_1G_3 &= &\left[\begin{array}{cc} 1 & 0 \\ 0 & -1 \end{array}\right]
\left[\begin{array}{cc} 0 & -1 \\ 1 & 0 \end{array}\right] \ = \ \left[\begin{array}{cc} 0 & -1 \\ -1 & 0 \end{array}\right]
\ = -\ G_2. \label{3.17} \ea
It follows from Eqs. (\ref{3.12}) -- (\ref{3.17}) that the generators satisfy the commutation relations
\be [G_1, G_2] = -2G_3, \ \ [G_2, G_3] = 2G_1, \ \ [G_3, G_1] = 2G_2, \label{3.18} \ee
where $[x, y] = xy - yx$. They also satisfy the anti-commutation relations
\be \{G_i, G_j\} = 0, \label{3.19} \ee
where $\{x, y\} = xy + yx$.

Solutions of Eq. (\ref{3.1}) are obtained by exponentiating $Gt$. Consider the generators separately. It follows from the identity $G_1^2 = I$ that
\ba \exp(G_1t) &= &\sum_n (G_1t)^n/n! \nonumber \\
&= &I + G_1t + It^2/2! + G_1t^3/3! \dots \nonumber \\
&= &I\cosh(t) + G_1\sinh(t) \nonumber \\
&= &\left[\begin{array}{cc} e^t & 0 \\ 0 & e^{-t} \end{array}\right]. \label{3.21} \ea
$G_2$ satisfies the same identity, so
\ba \exp(G_2t) &= &I\cosh(t) + G_2\sinh(t) \nonumber \\
&= &\left[\begin{array}{cc} C & S \\ S & C \end{array}\right], \label{3.22} \ea
where $C = \cosh(t)$ and $S = \sinh(t)$.
It follows from the identity $G_3^2 = -I$ that
\ba \exp(G_3t) &= &I + G_3t - It^2/2! - G_3t^3/3! \dots \nonumber \\
&= &I\cos(t) + G_3\sin(t) \nonumber \\
&= &\left[\begin{array}{cc} c & -s \\ s & c \end{array}\right], \label{3.23} \ea
where $c = \cos(t)$ and $s = \sin(t)$. Thus, $G_1$ generates dilations (stretches along the $x$-axis and squeezes along the $y$-axis), $G_2$ generates (Lorentz) boosts and $G_3$ generates (active) rotations. (It is for this reason that we chose $G_3 = -J$.)

Now consider the generators together. Let $G = G_1k_1 + G_2k_2 + G_3k_3$, where the coefficients $k_i$ include the time parameter. Then $G$ satisfies Eq. (\ref{3.4}) and its square
\ba G^2 &= &G_1^2k_1^2 + G_2^2k_2^2 + G_3^2k_3^2 + (G_1G_2 + G_2G_1)k_1k_2 \nonumber \\
&&+\ (G_2G_3 + G_3G_2)k_2k_3 + (G_3G_1 + G_1G_3)k_3k_1 \nonumber \\
&= &I(k_1^2 + k_2^2 - k_3^2), \label{3.24} \ea
because the generators anti-commute [Eq. (\ref{3.19})]. It follows from Eq. (\ref{3.24}) that
\ba \exp(G) &= &I + G + Ik^2/2! + Gk^2/3! + Ik^4/4! + \dots \nonumber \\
&= &I\cosh(k) + G\sinh(k)/k, \label{3.25} \ea
where $k = (k_1^2 + k_2^2 - k_3^2)^{1/2}$. Equation (\ref{3.25}) has the correct limits for dilation, boosting and rotation [Eqs. (\ref{3.21}) -- (\ref{3.23})]. Furthermore, by comparing Eqs. (\ref{2.6}) and (\ref{3.25}), one finds that the symmetric dilation matrix
\be D(\la, \th) = IC + G_1Sk_1/k + G_2Sk_2/k, \label{3.26} \ee
where $C = \cosh(k)$, $S = \sinh(k)$, $k = (k_1^2 + k_2^2)^{1/2}$, $\la = k$ and $\tan(2\th) = k_2/k_1$.
It follows from the properties of the exponential function that the inverse matrix
\be \exp(-G) = I\cosh(k) - G\sinh(k)/k. \label{3.27} \ee
One can verify that the matrices in Eqs. (\ref{3.25}) and (\ref{3.27}) are inverses by multiplying them and using the identity $G^2 = Ik^2$, or by using Eq. (\ref{2.2}) and the identities $J^tG_i^tJ = -G_i$.

The preceding derivation of Eq. (\ref{3.25}) depends on the anti-commutation relations (\ref{3.19}). There is an alternative derivation that does not depend on these relations \cite{gil08}. Let $G$ be a $2 \times 2$ matrix with eigenvalues $\la_1$ and $\la_2$. The Cayley--Hamilton (CH) theorem states that $G$ satisfies its own characteristic equation. Hence,
\be G^2 = -\la_1\la_2I + (\la_1 + \la_2)G. \label{3.31} \ee
Now let $G$ be the generating matrix $G_1k_1 + G_2k_2 + G_3k_3$. Then, written explicitly,
\be G = \left[\begin{array}{cc} k_1 & k_2 - k_3 \\ k_2 + k_3 & -k_1 \end{array}\right]. \label{3.32} \ee
It is easy to verify that the eigenvalues of $G$ are $\pm k$, from which it follows that $\la_1 + \la_2 = 0$ and $-\la_1\la_2 = k^2$. Hence, $G^2 = k^2I$, as stated above.
The eigenvectors of $G$ need not be orthogonal, because $G$ is not necessarily symmetric.
Nonetheless, $G$ and $M = \exp(G)$ have the same eigenvectors, and $\la_m = e^{\la_g} = e^{\pm k}$. This property is discussed further in App. A.

If the generating parameters $k_1$, $k_2$ and $k_3$ are specified, then it follows from Eqs. (\ref{3.25}) and (\ref{3.32}) that
\be M = \left[\begin{array}{cc} C_0 + S_1 & S_2 - S_3 \\ S_2 + S_3 & C_0 - S_1 \end{array}\right], \label{3.41} \ee
where $C_0 = \cosh(k)$, $S_0 = \sinh(k)$, $S_i = S_0k_i/k$ and $k = (k_1^2 + k_2^2 - k_3^2)^{1/2}$. It is easy to verify that $M$ has determinant 1 and eigenvalues $C_0 \pm S_0$.
Conversely, if $M$ is specified, can these parameters be deduced? By comparing the matrices in Eqs. (\ref{2.32}) and (\ref{3.41}), one finds that
\be C_0 = (\al + \de)/2, \ S_1 = (\al - \de)/2, \ S_2 = (\bt + \ga)/2, \ S_3 = (\ga - \bt)/2. \label{3.42} \ee
[These formulas are equivalent to the ones for $\al'$, $\de'$, $\bt'$ and $\ga'$, which were stated in Eqs. (\ref{2.33}).]
With $C_0$, $S_1$, $S_2$ and $S_3$ known, $S_0 = (C_0^2 - 1)^{1/2}$, or $(S_1^2 + S_2^2 - S_3^2)^{1/2}$, and
\be k = \log(C_0 + S_0). \label{3.43} \ee
With $S_0$ and $k$ known,
\be k_1/k = S_1/S_0, \ \ k_2/k = S_2/S_0, \ \ k_3/k = S_3/S_0. \label{3.44} \ee
Equations (\ref{3.42}) -- (\ref{3.44}) specify the generating parameters $k_1$, $k_2$ and $k_3$ as functions of the components $\al$, $\bt$, $\ga$ and $\de$. They can also be specified as functions of the parameters $\la$, $\th_1$ and $\th_2$. By comparing Eqs. (\ref{2.31}) and (\ref{3.41}), one finds that
\be C_0 = Cc_-, \ S_1 = Sc_+, \ S_2 = Ss_+, \ S_3 = Cs_-, \label{3.45} \ee
where $C = \cosh(\la)$, $c_\pm = \cos(\th_\pm)$  and $\th_\pm = \th_2 \pm \th_1$.
The definitions of $S$ and $s_\pm$ are similar.
Equations (\ref{3.43}) and (\ref{3.44}) still apply.
The preceding formulas are valid for arbitrary values of $\al$, $\bt$, $\ga$ and $\de$, or $\la$, $\th_1$ and $\th_2$. Hence, there is a one-to-one relationship between symplectic matrices and their generators.

In Sec. 2, we showed that $M = R_2DR_1^t = R_{21}N_1 = N_2R_{21}$, where $R_{21} = R_2R_1^t$ is a rotation matrix, and $N_1 = R_1DR_1^t$ and $N_2 = R_2DR_2^t$ are dilation matrices [Eqs. (\ref{2.5}) and (\ref{2.6})]. It follows from Eq. (\ref{3.23}) that
\be R_{21} = \exp(G_3l_3), \label{3.51} \ee
where $l_3 = \th_{21} = \th_-$. It follows from Eq. (\ref{3.26}) that
\be N_1 = \exp(G_1l_1 + G_2l_2), \label{3.52} \ee
where  $l = (\l_1^2 + l_2^2)^{1/2} = \la$, $l_1 = l\cos(2\th_1)$ and $l_2 = l\sin(2\th_1)$. Likewise,
\be N_2 = \exp(G_1l_1' + G_2l_2'), \label{3.53} \ee
where  $l' = [(\l_1')^2 + (l_2')^2]^{1/2} = \la$, $l_1' = l\cos(2\th_2)$ and $l_2' = l\sin(2\th_2)$. Notice that $l' = l$, because $N_1$ and $N_2$ are related by a similarity transform. (They have the same dilation parameter, but different orientations).

In the preceding paragraphs, we specified the $k$- and $l$-parameters as functions of $\la$, $\th_1$ and $\th_2$.
It is useful to specify the $l$-parameters in terms of the $k$-parameters directly. Suppose that $k_1$ -- $k_3$, and, hence, $C_0$ and $S_1$ -- $S_3$ are known. Then it follows from Eqs. (\ref{3.45}) that $C = (C_0^2 + S_3^2)^{1/2}$, $S = (S_1^2 + S_2^2)^{1/2}$ and $\la = \log(C + S)$.
Alternatively, by using the identity $k_1^2 + k_2^2 = k^2 + k_3^2$, one can rewrite the second equation in the form
\be \sinh(\la) = (S_1^2 + S_2^2)^{1/2} = (1 + k_3^2/k^2)^{1/2}\sinh(k). \label{3.54b} \ee
Notice that $\la = l$ only differs from $k$ if $k_3$ is nonzero.
It also follows from Eqs. (\ref{3.45}) that
\be \tan(\th_+) = S_2/S_1 = k_2/k_1, \ \ \tan(\th_-) = S_3/C_0 = (k_3/k)\tanh(k). \label{3.54} \ee
Notice that $\th_+$ is constant and $\th_-$ is only nonzero if $k_3$ is nonzero. By using identities (\ref{2.40}), one finds that
\be \tan(2\th_1) = {C_0S_2 - S_1S_3 \over C_0S_1 + S_2S_3}, \ \ \tan(2\th_2) = {C_0S_2 + S_1S_3 \over C_0S_1 - S_2S_3}. \label{3.55}\ee
[Equations (\ref{3.54}) are equivalent to Eqs. (\ref{2.38}) and (\ref{2.39}), and Eqs. (\ref{3.55}) are equivalent to Eqs. (\ref{2.41}) and (\ref{2.42}).]
The related cosines and sines follow from the identities $c = 1/(1 + t^2)^{1/2}$ and $s = t/(1 + t^2)^{1/2}$. 
By combining the preceding results, one finds that
\ba &&l_1 = (C_0S_1 + S_2S_3)l/CS,  \ \ l_2 = (C_0S_2 - S_1S_3)l/CS, \label{3.56} \\
&&l_1' = (C_0S_1 - S_2S_3)l/CS,  \ \ l_2' = (C_0S_2 + S_1S_3)l/CS, \label{3.57} \ea
where $C$, $S$ and $l$ are functions of $C_0$ and $S_1$ -- $S_3$. In both cases, $\tan(l_3) = S_3/C_0$. Notice that one can convert the first of Eqs. (\ref{3.55}) to the second, and Eqs. (\ref{3.56}) to Eqs. (\ref{3.57}), by changing the sign of $S_3$. This result is consistent with the second of Eqs. (\ref{3.54}). By combining Eqs. (\ref{3.45}) with Eqs. (\ref{3.56}) and (\ref{3.57}), one can show that the latter equations are consistent with the simpler equations for the $l$-parameters that were stated after Eqs. (\ref{3.52}) and (\ref{3.53}).

In applied mathematics and quantum optics, formulas (\ref{3.56}) and (\ref{3.57}) are called decomposition (disentangling) formulas, because they allow $\exp(G_1k_1 + G_2k_2 + G_3k_3)$, in which the generators appear symmetrically (together), to be rewritten as $\exp(G_3l_3)\exp(G_1l_1 + G_2l_2)$ or $\exp(G_1l_1' + G_2l_2')\exp(G_3l_3)$, in which $G_1$ and $G_2$, and $G_3$, appear asymmetrically (separately). It is also possible to disentangle the three generators completely \cite{wod85,ger01}.

\newpage

\sec{4. Matrix products}

In this section, we discuss matrix products. Let $M_1$ and $M_2$ be two symplectic matrices, and let $M_3 = M_2M_1$ be their product. Then, in the notation of Eq. (\ref{2.32}),
\ba \left[\begin{array}{cc} \al_3 & \bt_3 \\ \ga_3 & \de_3 \end{array}\right]
&= &\left[\begin{array}{cc} \al_2 & \bt_2 \\ \ga_2 & \de_2 \end{array}\right]
\left[\begin{array}{cc} \al_1 & \bt_1 \\ \ga_1 & \de_1 \end{array}\right] \nonumber \\
&= &\left[\begin{array}{cc} \al_2\al_1 + \bt_2\ga_1 & \al_2\bt_1 + \bt_2\de_1 \\ \ga_2\al_1 + \de_2\ga_1 & \ga_2\bt_1 + \de_2\de_1 \end{array}\right]. \label{4.1} \ea
Equation (\ref{4.1}) specifies the product matrix in terms of components. One can rewrite it in terms of dilation and angle parameters by using Eqs. (\ref{2.37}), (\ref{2.41}) and (\ref{2.42}).

If the constituent matrices are symmetric ($\bt_i = \ga_i$), then the product matrix is symmetric if and only if $\al_2\bt_1 + \bt_2\de_1 = \bt_2\al_1 + \de_2\bt_1$. This condition might, or might not, be satisfied. Thus, the composition of two dilations is not necessarily a dilation. According to Eqs. (\ref{2.12}) and (\ref{2.14}), it is a dilation, which is specified by Eqs. (\ref{2.37}) and (\ref{2.41}), followed by a rotation, which is specified by Eq. (\ref{2.39}).

Alternatively, in the notation of Eq. (\ref{3.41}),
\be \left[\begin{array}{cc} C_{30} + S_{31} & S_{32} - S_{33} \\ S_{32} + S_{33} & C_{30} - S_{31} \end{array}\right]
= \left[\begin{array}{cc} C_{20} + S_{21} & S_{22} - S_{23} \\ S_{22} + S_{23} & C_{20} - S_{21} \end{array}\right]
\left[\begin{array}{cc} C_{10} + S_{11} & S_{12} - S_{13} \\ S_{12} + S_{13} & C_{10} - S_{11} \end{array}\right]. \ee
It is easy to verify that
\ba C_{30} &= &C_{20}C_{10} + S_{21}S_{11} + S_{22}S_{12} - S_{23}S_{13}, \label{4.2} \\
S_{31} &= &C_{20}S_{11} + S_{21}C_{10} + S_{22}S_{13} - S_{23}S_{12}, \label{4.3} \\
S_{32} &= &C_{20}S_{12} - S_{21}S_{13} + S_{22}C_{10} + S_{23}S_{11}, \label{4.4} \\
S_{33} &= &C_{20}S_{13} - S_{21}S_{12} + S_{22}S_{11} + S_{23}C_{10}. \label{4.5} \ea
Equations (\ref{4.2}) -- (\ref{4.5}), together with Eqs. (\ref{3.43}) and (\ref{3.44}), specify the product matrix in terms of generator coefficients.
The symmetry condition $S_{33} = 0$ is not necessarily satisfied, even if $S_{23} = S_{13} = 0$.

Although the component and generator forms of the product matrix and symmetry condition are precise, they provide no physical insight.
In the dilation--angle representation of dilations ($M_i = R_iD_iR_i^t$), the composite transformation
\be Y = (R_2D_2R_2^t)(R_1D_1R_1^t)X. \label{4.6} \ee
In Eq. (\ref{4.6}), the angles are measured relative to the $x$-axis. If they are measured relative to the $x'$-axis, which makes the angle $\th_1$ with the $x$-axis, then
\be (R_1^tY) = (R_2R_1^t)D_2(R_2^tR_1)D_1(R_1^tX). \label{4.7} \ee
In the rotated coordinate system, the composite transformation depends on the difference angle $\th_{21} = \th_2 - \th_1$. Let $M_3 = R_{21}D_2R_{21}^tD_1$. Then, written explicitly,
\ba M_3 &= &\left[\begin{array}{cc} c(C_2 + S_2) & -s(C_2 - S_2) \\ s(C_2 + S_2) & c(C_2 - S_2) \end{array}\right]\left[\begin{array}{cc} c(C_1 + S_1) & s(C_1 - S_1) \\ -s(C_1 + S_1) & c(C_1 - S_1) \end{array}\right] \\
&= &\left[\begin{array}{cc} C_2C_1 + C_2S_1 + (S_2S_1 + S_2C_1)c_{21}  &  (S_2C_1 - S_2S_1)s_{21} \\ (S_2C_1 + S_2S_1)s_{21} & C_2C_1 -  C_2S_1 + (S_2S_1 - S_2C_1)c_{21} \end{array}\right], \nonumber \label{4.8} \ea
where $C_i = \cosh(\la_i)$, $c = \cos(\th_{21})$ and $c_{21} = \cos(2\th_{21})$.
Notice that the second cosine involves a double angle.
The definitions of $S_i$, $s$ and $s_{21}$ are similar.
For nontrivial dilations ($\la_i \neq 0$), the symmetry condition is $s_{21} = 0$, which means that $\th_{21} = 0$ (parallel dilations) or $\pi/2$ (perpendicular dilations).

In the notation of Eq. (\ref{2.32}),
\be \al_3' = C_2C_1 + S_2S_1c_{21}, \ \ \bt_3' = S_2C_1s_{21}, \ \ 
\ga_3' = S_2S_1s_{21}, \ \ \de_3' = C_2S_1 + S_2C_1c_{21}. \label{4.9} \ee
According to Eqs. (\ref{2.35}) and (\ref{2.36}),
\ba C_3^2 &= &(\al_3')^2 + (\ga_3')^2 \ = \ C_2^2C_1^2 + S_2^2S_1^2 + 2C_2C_1S_2S_1c_{21}, \label{4.10} \\
S_3^2 &= &(\bt_3')^2 + (\de_3')^2 \ = \ C_2^2S_1^2 + C_1^2S_2^2 + 2C_2S_1C_1S_2c_{21}. \label{4.11} \ea
With $C_3$ and $S_3 \ge 0$ known, the dilation parameter $\la_3 = \log(C_3 + S_3)$. The assumption that $S_3$ is non-negative is equivalent to the assumption that the first Schmidt coefficient, $C_3 + S_3$, is larger than the second coefficient, $C_3 - S_3$.

The exponential dilation parameter, $E_3 = C_3 + S_3$, is plotted as a function of the difference angle in Fig. 1, for cases in which $\la_1 \ge \la_2$. For $\th_{21} = 0$, $C_3 = C_2C_1 + S_2S_1 = \cosh(\la_1 + \la_2)$ and $S_3 = S_2C_1 + C_2S_1 = \sinh(\la_1 + \la_2)$, so the composite dilation is the product of the constituent ones ($E_3 = E_2E_1$). In contrast, for $\th_{21} = \pi/2$, $C_3 = C_2C_1 - S_2S_1 = \cosh(\la_1 - \la_2)$ and $S_3 = S_1C_2 - C_1S_2 = \sinh(\la_1 - \la_2)$, so the composite dilation is the ratio of the constituent ones ($E_3 = E_1/E_2$). In particular, if $E_2 = E_1$, then the second transformation is the inverse of the first. This result makes sense, because for perpendicular transformations, the roles of the stretching and squeezing axes are interchanged.
\begin{figure}[h!]
\vspace*{0.1in} 
\centerline{\includegraphics[width=3in]{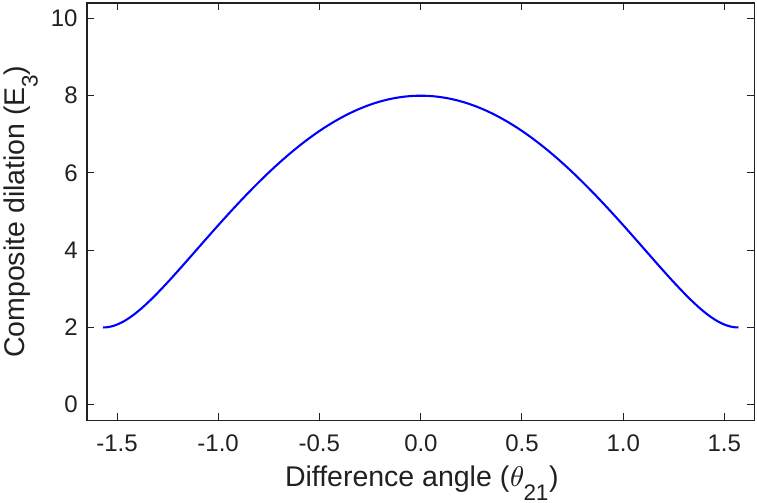} \hspace*{0.2in} \includegraphics[width=3in]{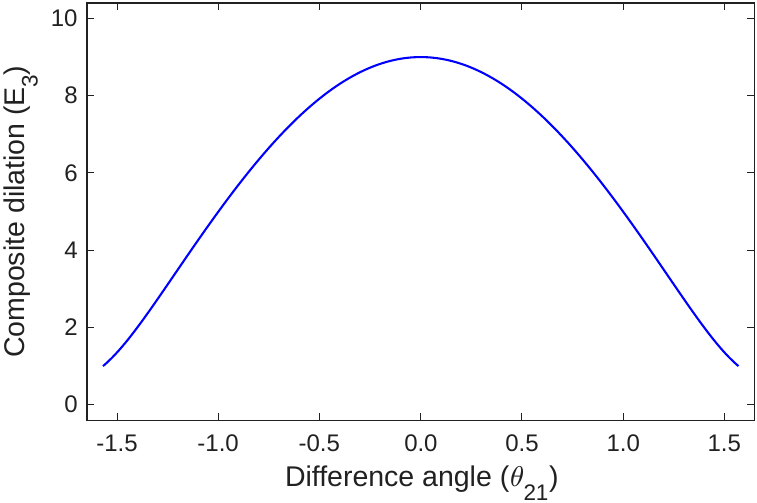}}
\vspace*{-0.1in} 
\caption{Composite dilation $E_3$ plotted as a function of the difference angle $\th_{12}$, for $E_1 = 4$ and $E_2 = 2$ (left), and $E_1 = E_2 = 3$ (right). These results are based on Eqs. (\ref{4.10}) and (\ref{4.11}).}
\end{figure}

The product matrix is not necessarily symmetric, so $M_3 = Q_3D_3P_3^t$, where $P_3 = R(\ph_3)$ and $Q_3 = R(\th_3)$. According to Eqs. (\ref{2.38}) and (\ref{2.39}),
\be \tan(\th_3 + \ph_3) = {C_1S_2s_{21} \over C_2S_1 + C_1S_2c_{21}}, \ \ 
\tan(\th_3 - \ph_3) = {S_2S_1s_{21} \over C_2C_1 + S_2S_1c_{21}}. \label{4.12} \ee
By combining Eqs. (\ref{4.12}) with identity (\ref{2.40}), one finds that
\be \tan(2\ph_3) = {T_2s_{21} \over D_2T_1 + T_2D_1c_{21}}, \ \
\tan(2\th_3) = {T_2D_1s_{21} + S_2^2T_1t_{21} \over C_2^2T_1 + T_2D_1c_{21} + S_2^2T_1d_{21}}, \label{4.13} \ee
where $D_i = \cosh(2\la_i)$, $T_i = \sinh(2\la_i)$, $d_{21} = \cos(4\th_{21})$ and $t_{21} = \sin(4\th_{21})$.
The second of Eqs. (\ref{4.13}) is equivalent to the equation
%
\be \tan[2(\th_3 - \th_{21})] = {-T_1s_{21} \over T_2D_1 + D_2T_1c_{21}}, \label{4.14} \ee
which does not involve quadruple angles (App. C).

The composite angles are plotted as functions of the difference angle in Figs. 2 and~3.
[We used Eqs. (\ref{4.12}) to make both figures, to avoid the arctangent issues associated with Eqs. (\ref{4.13}) and (\ref{4.14}).]
For $\th_{21} = 0$, the input and output angles $\ph_3 = \th_3 = 0$: The product of two parallel dilations is another parallel dilation. For $\th_{21} = \pm \pi/2$, the situation is more complicated. For the general case in which $E_1 \neq E_2$, $\ph_3 = \th_3 = 0$: The product of two perpendicular dilations is a parallel dilation. Because the roles of the stretching and squeezing axes are interchanged, the net result is stretching and squeezing along the same pair of axes, with magnitudes to be determined. For the special case in which $E_1 = E_2$, the second transformation is the inverse of the first. Nonetheless, $\ph_3 = \th_3 = \pm \pi/4$: The product of two perpendicular dilations appears to be a boost (a dilation at $\pi/4$ radians to the $x'$ axis). However, for this case, $C_3 = 1$ and $S_3 = 0$. A boost of zero magnitude is the identity transformation. This strange
\begin{figure}[h!]
\vspace*{0.0in} 
\centerline{\includegraphics[width=3in]{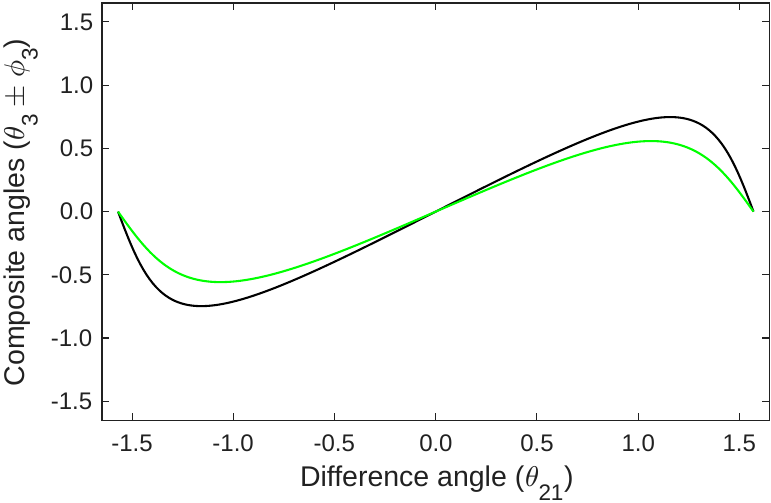} \hspace*{0.2in} \includegraphics[width=3in]{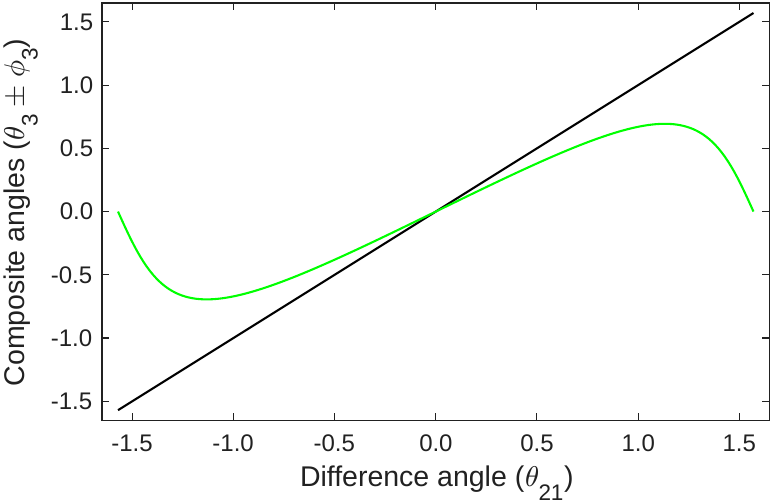}}
\vspace*{-0.1in} 
\caption{Composite sum angle (black) and difference angle (green), plotted as functions of the difference angle $\th_{21}$, for $E_1 = 4$ and $E_2 = 2$ (left), and $E_1 = E_2 = 3$ (right). These results are based on Eqs. (\ref{4.12}).}
\end{figure}
\begin{figure}[h!]
\vspace*{0.0in} 
\centerline{\includegraphics[width=3in]{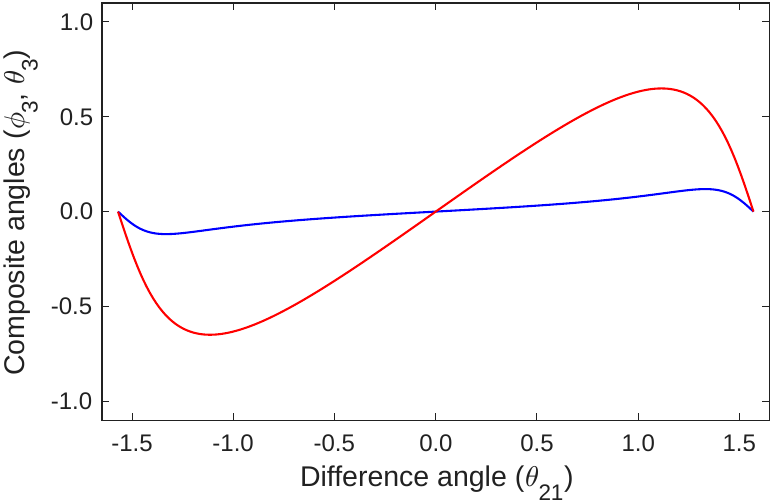} \hspace*{0.2in} \includegraphics[width=3in]{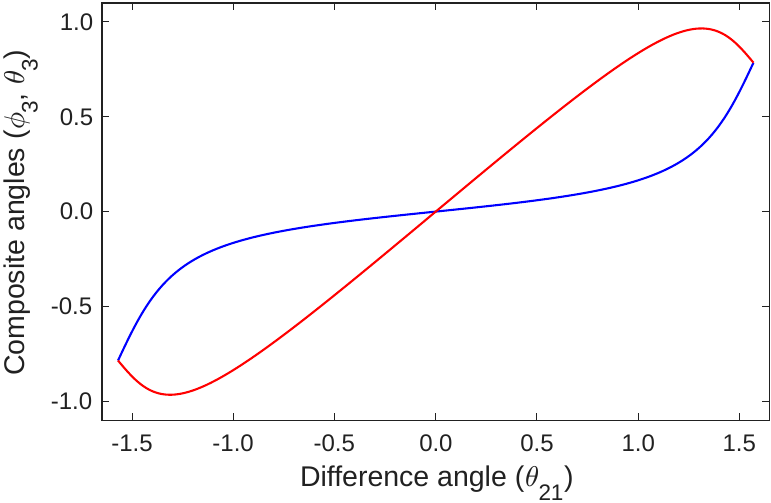}}
\vspace*{-0.1in} 
\caption{Composite input angle (blue) and output angle (red), plotted as functions of the difference angle $\th_{21}$, for $E_1 = 4$ and $E_2 = 2$ (left), and $E_1 = E_2 = 3$ (right). These results are based on Eqs.~(\ref{4.12}).}
\end{figure}

\noindent result is possible because the Schmidt decomposition of the identity transformation is not unique: $I = RIR^t$ for any rotation matrix $R$.
The products of asymmetric (dilation--rotation) matrices are discussed in App. C.

\vspace*{3in}

\newpage

\sec{5. Matrix optics}

In this section, we discuss ray-tracing matrices briefly \cite{ger94,pea15}, to illustrate the occurrence and use of symplectic matrices, generators and decompositions.
An optical ray is a pencil of light. Let $z$ be the distance along the axis of an optical system and $x$ be the distance from the axis. Then at any point in space, the ray is characterized by its position ($x$) and slope ($\th = dx/dz$). In general, the ray is curved, because the refractive index $n$ varies with position ($x$ and $z$). One can think of the ray as the path taken by an imaginary corpuscle of light as it moves through the system.

By using the principles of geometric optics (reflection and refraction at a dielectric interface), one can determine the effects on the ray of a variety of optical elements, such as distances (spaces), lenses, mirrors and prisms.
In the paraxial (small-angle) approximation, the effect of an optical element on a ray is described by the input--output equation
\be \left[\begin{array}{c} x_1 \\ n_1\th_1 \end{array}\right] = \left[\begin{array}{cc} a & b \\ c & d \end{array}\right]
\left[\begin{array}{c} x_0 \\ n_0\th_0 \end{array}\right], \label{y1} \ee
where $a$ -- $d$ are real parameters and the determinant $ad - bc$ equals 1 (because \'etendue is conserved). In general, the element changes the position and slope of the ray. Notice that $x$ and $\th$ have different units, so $b$ and $c$ are dimensional.
The matrix in Eq. (\ref{y1}) is called a ray-transfer matrix. Such matrices are invertible because their determinants are nonzero. Consequently, the set of transfer matrices forms a group under multiplication, namely the special linear group SL(2), which equals the symplectic group Sp(2).

It is instructive to consider some examples (in which $n_1 = n_0 = 1$). For propagation in space, the transfer matrix
\be M = \left[\begin{array}{cc} 1 & z \\ 0 & 1 \end{array}\right]. \label{y2} \ee
The displacement of a ray that is inclined to the optical axis increases linearly with distance.
It is easy to verify that $M(z_2)M(z_1) = M(z_2 + z_1)$. Propagation through the distance $z_1$ followed by propagation through the distance $z_2$ is equivalent to propagation through the total distance $z_2 + z_1$. This result is sensible.

For transmission through a thin lens, the transfer matrix
\be M = \left[\begin{array}{cc} 1 & 0 \\ -1/f & 1 \end{array}\right], \label{y3} \ee
where $f$ is the focal length.
For a convex (focusing) lens, $f$ is positive. A horizontal incident ray moves toward the axis and intersects it at a distance $f$ from the lens. In contrast, for a concave (defocusing) lens, $f$ is negative and a horizontal incident ray moves away from the axis.
It is easy to verify that $M(1/f_2)M(1/f_1) = M(1/f_2 + 1/f_1)$. The inverse focal length of the composite device is the sum of the individual inverse lengths. This relation is called the lens law.

Some optical systems provide magnification, for which the putative transfer matrix
\be M = \left[\begin{array}{cc} m & 0 \\ 0 & 1/m \end{array}\right]. \label{y4} \ee
The determinant condition ensures that the magnification factors ($a = m$ and $d = 1/m$) are inverses. If the image is larger than the object, then the output slope is smaller than the input slope.

Consider an imaging system with two distances and a lens, for which the composite transfer matrix
\ba   M &= &\left[\begin{array}{cc} 1 & z_2 \\ 0 & 1 \end{array}\right]
\left[\begin{array}{cc} 1 & 0 \\ -1/f & 1 \end{array}\right]
\left[\begin{array}{cc} 1 & z_1 \\ 0 & 1 \end{array}\right] \nonumber \\
&=  &\left[\begin{array}{cc} 1 - z_2/f & z_1 + z_2(1 - z_1/f) \\ -1/f & 1 - z_1/f \end{array}\right]. \label{y5} \ea
The imaging condition is that the output position is independent of the input slope ($b = 0$), which means that every ray emitted from an object point arrives at the same image point. This condition can be written in the form $1/z_1 + 1/z_2 = 1/f$. Suppose, for simplicity, that $f$ is positive. Then $z_1$ and $z_2$ must be larger than $f$. If the imaging condition is satisfied, then $a = -z_2/z_1$,  $d = -z_1/z_2$ and the transfer matrix
\be M = \left[\begin{array}{cc} -z_2/z_1 & 0 \\ -1/f & -z_1/z_2 \end{array}\right]. \label{y6} \ee
Negative magnification factors correspond to an inverted image. Matrix (\ref{y6}) is similar, but not identical, to matrix (\ref{y4}).

If one places a second lens after the second distance, then the composite transfer matrix
\ba M &= &\left[\begin{array}{cc} 1 & 0 \\ -1/f_2 & 1 \end{array}\right]
\left[\begin{array}{cc} -z_2/z_1 & 0 \\ -1/f_1 & -z_1/z_2 \end{array}\right] \nonumber \\
&= &\left[\begin{array}{cc} -z_2/z_1 & 0 \\ z_2/z_1f_2 - 1/f_1 & -z_1/z_2 \end{array}\right]. \label{y7} \ea
If $f_2 = (z_2/z_1)f_1$, then $c = 0$. The slope of the output ray is independent of the position of the input ray. In this case, matrix (\ref{y7}) is identical to matrix (\ref{y4}). Thus, the canonical matrix is associated with a modest system of two distances and two lenses.

One obtains a similar result by using two lenses ($f_1$ and $f_2$) separated by the distance $z_1 = f_1 + f_2$. (This device is called a beam expander.) If one moves the image plane a distance $z_2 = (f_2/f_1)z_1$ behind the second lens, then $b = 0$ and the transfer matrix takes the canonical form.

By using the identity $ad - bc = 1$, one can verify that
\ba \left[\begin{array}{cc} a & b \\ c & d \end{array}\right] &= &\left[\begin{array}{cc} 1 & 0 \\ c/a & 1 \end{array}\right]
\left[\begin{array}{cc} a & 0 \\ 0 & 1/a \end{array}\right]
\left[\begin{array}{cc} 1 & b/a \\ 0 & 1 \end{array}\right] \nonumber \\
&= &\left[\begin{array}{cc} 1 & b/d \\ 0 & 1 \end{array}\right]
\left[\begin{array}{cc} 1/d & 0 \\ 0 & d \end{array}\right]
\left[\begin{array}{cc} 1 & 0 \\ c/d & 1 \end{array}\right]. \label{y8} \ea
Every ray transformation can be decomposed as propagation followed by magnification and focusing, or focusing followed by magnification and propagation. This well-known result implies that every transformation can be effected by a small number of optical elements.

It is convenient to denote the matrices in Eqs. (\ref{y2}), (\ref{y3}) and (\ref{y4}) by $M_2$, $M_3$ and $M_1$, respectively. These matrices have the generators
\be G_1 = \left[\begin{array}{cc} 1 & 0 \\ 0 & -1 \end{array}\right], \ \ 
G_2 = \left[\begin{array}{cc} 0 & 1 \\ 0 & 0 \end{array}\right], \ \ 
G_3 = \left[\begin{array}{cc} 0 & 0 \\ -1 & 0 \end{array}\right]. \label{y9} \ee
To be precise, $M_1 = e^{G_1\log m}$, $M_2 = e^{G_2z}$ and $M_3 = e^{G_3/f}$.
Notice that $G_1$, which produces a dilation, is the first generator of Sp(2), whereas $G_2$ and $G_3$, which produce shears, are linear combinations of the second and third generators, so generators (\ref{y9}) are alternative generators of the same group. They are preferable, in the context of ray optics, because they correspond to fundamental operations (magnification, propagation and focusing).

Generators (\ref{y9}) satisfy the commutation relations
\be [G_1, G_2] = 2G_2, \ \ [G_2, G_3] = -G_1, \ \ [G_3, G_1] = 2G_3. \label{y10} \ee
The modified generators $K_0 = G_1/2$, $K_+ = G_2$ and $K_- = G_3$ satisfy the modified relations
\be [K_0, K_\pm] = \pm K_\pm, \ \ [K_+, K_-] = -2K_0. \label{y11} \ee
Relations (\ref{y11}) also appear in the quantum theory of photon generation by spontaneous three- or four-wave mixing, which is governed by the group SU(1,1) \cite{wod85,ger01}. This appearance indicates that Sp(2) has the same structure as SU(1,1). We will discuss this relation further in \cite{mck24b}.

\newpage

\sec{6. Summary}

In this article, we reviewed the properties of the symplectic group Sp(2), which is the set of real $2 \times 2$ matrices with determinant 1 [Eq. (\ref{2.3})].
A symplectic matrix $M$ is specified by four real parameters, $\al$, $\bt$, $\ga$ and $\de$, only three of which are independent.
Alternatively, it can be specified by a dilation parameter, $\la$, and two rotation parameters, $\th_1$ and $\th_2$ [Eq.~(\ref{2.31})].

In general, symplectic matrices are asymmetric. A symmetric matrix corresponds to a dilation with axes that are inclined relative to the coordinate axes [Eq. (\ref{2.6})]. An asymmetric matrix can be written as the product of a symmetric matrix and a rotation matrix, in either order: $M = R_{21}N_1 = N_2R_{21}$ [Eqs. (\ref{2.11}) and (\ref{2.12}), and the text after Eq. (\ref{2.14})].

Every real matrix has the Schmidt decomposition $M = QDP^t$, where $D$ is a diagonal matrix, and $P$ and $Q$ are orthogonal matrices. The entries of $D$, which are called Schmidt coefficients, are the dilation (stretching and squeezing) parameters of the transformation. The columns of $P$ and $Q$, which are called input and output vectors, determine the principal axes of the dilation: $P$ specifies the axes of $N_1$, whereas $Q$ specifies the axes of $N_2$. The product $QP^t$ specifies the rotation matrix $R_{21}$ [Eq. (\ref{2.31}) and the text after Eq. (\ref{2.42})].

Every symplectic matrix, with determinant 1, can be written as the exponential of a generating matrix, with trace 0. We derived formulas for the matrix components in terms of the generator coefficients [Eq. (\ref{3.41})]. We also derived formulas for the generator coefficients in terms of the components [Eqs. (\ref{3.42}) -- (\ref{3.44})], and the dilation and angle parameters [Eqs. (\ref{3.43}) -- (\ref{3.45})]. We derived formulas for the generator coefficients of the factor matrices, $N_1$, $N_2$ and $R_{21}$, in terms of the dilation and angle parameters [Eqs. (\ref{3.51}) -- (\ref{3.53})]. We also related the generator coefficients of the factor matrices to the generator coefficients of the matrix directly [Eqs. (\ref{3.56}) and (\ref{3.57})]. These formulas are examples of disentanglement formulas, which pervade applied mathematics and quantum optics \cite{wod85,ger01}.

The composition of two dilations is not (usually) another dilation. Rather, it is a dilation followed by a rotation (as required by the Schmidt decomposition theorem). We derived formulas for the dilation and angle parameters of the product matrix in terms of the parameters of the constituent matrices [Eqs. (\ref{4.10}), (\ref{4.11}), (\ref{4.13}) and (\ref{4.14})]. The composite dilation parameter depends on the constituent dilation parameters and the difference between the constituent angles. For parallel dilations, the dilation parameters add, whereas for perpendicular dilations, they subtract.

In Sec. 5, we discussed ray-tracing matrices. We stated the transfer matrices for propagation and focusing [Eqs. (\ref{y2}) and (\ref{y3})], and described two equivalent ways to produce ideal magnification [Eq. (\ref{y7}) and the text that follows]. Every ray transformation can be decomposed into propagation, magnification and focusing [Eq. (\ref{y8})]. In classical dynamics, the natural generators correspond to dilation, boosting and rotation [Eqs. (\ref{3.11})], whereas in ray optics, they correspond to dilation, propagation and focusing [Eqs. (\ref{y9})]. Nonetheless, the two sets of generators are equivalent, because they are linear combinations of each other.

In App. A, we discuss the adjoint decomposition of an arbitrary real matrix, which is a generalization of the spectral decomposition of a symmetric real matrix. In App. B, we discuss the Schmidt decomposition of a symplectic matrix. Both appendices contain specific formulas for $2 \times 2$ matrices and some comments on higher-order matrices. In App. C, we discuss the Schmidt decomposition of the product of two asymmetric (dilation--rotation) matrices.

Many of the results derived herein for matrices in Sp(2) have analogs for matrices in the special unitary group SU(1,1), which governs photon-pair generation, and the special orthogonal group SO(1,2), which governs Lorentz transformation in time and two space dimensions \cite{kim83,mck24a}. Deriving and illustrating these results for $2 \times 2$ real matrices facilitates the derivation and comprehension of similar, but more complicated, results for $2 \times 2$ complex and $3 \times 3$ real matrices. We will describe the relations among Sp(2), SU(1,1) and SO(1,2) in a future article \cite{mck24b}.

\newpage

\sec{Appendix A: Adjoint decomposition}

In this appendix, we discuss the adjoint decomposition of a real matrix, which need not be symmetric. Let $M$ be an $n \times n$ matrix. Then $M$ and its transpose $M^t$ share the same eigenvalues $\la_i$. Now let $E_i$ and $F_i$ be the corresponding eigenvectors of $M$ and $M^t$, respectively. Then
\be \la_iF_j^tE_i = F_j^tME_i = E_i^tM^tF_j = \la_j E_i^tF_j, \label{a1} \ee
because the inner products are scalars. Hence,
\be (\la_j - \la_i)F_j^tE_i = 0. \label{a2} \ee
The eigenvectors $E_i$ and their adjoints $F_j$ are orthogonal ($i \neq j$). By normalizing them appropriately, one can can make them orthonormal ($F_j^tE_i = \de_{ij}$). Let $E$ be the matrix whose $i$th column is $E_i$ and $F$ be the matrix whose $j$th column is $F_j$. Then it follows from the orthonormality equation that $F^tE = I = E^tF$. If the eigenvectors of $M$ form a complete set, then any vector $X$ can be written as the linear combination $\tsum_i x_iE_i$, where $x_i = F_i^tX$. Hence, $X = \tsum_i E_iF_i^t X$. By writing the outer products in terms of components, it is easy to verify that $\tsum_i E_iF_i^t = EF^t$, from which it follows that $EF^t = I$. Now let $\La = \diag(\la_i)$. Then it is easy to verify that the matrices $M$ and $E\La F^t$ have the same effect on an arbitrary vector $X$, so they are equal. The adjoint decomposition $M = E\La F^t$, where $M$ is asymmetric, is a generalization of the spectral decomposition $N = E\La E^t$, where $N$ is symmetric. [In passing, if $M$ is complex, then $M$ and $M^\d$ have conjugate eigenvalues and the adjoint decomposition is $M = E\La F^\d$.]

Two examples are relevant. The first concerns the $2 \times 2$ matrix
\be M = \left[\begin{array}{cc} \al & \bt \\ \ga & \de \end{array}\right], \label{a3} \ee
which has the eigenvalues
\be \la_\pm = (\al + \de)/2 \pm [(\al + \de)^2/4 - (\al\de - \bt\ga)]^{1/2}. \label{a4} \ee
If the matrix is symplectic, then $\al\de - \bt\ga = 1$. For this case, the squared eigenvalues
\ba \la_\pm^2 &= &2(\al')^2 - 1 \pm 2\al'[(\al')^2 - 1]^{1/2} \nonumber \\
&= &2(\al')^2 - 1 \pm [4(\al')^4 - 4(\al')^2]^{1/2} \nonumber \\
&= &2(\al')^2 - 1 \pm \{[2(\al')^2 - 1]^2 - 1\}^{1/2}, \label{a5} \ea
where $\al' = (\al + \de)/2$ and the recurring term
\ba 2(\al')^2 - 1 &= &(\al^2 + 2\al\de + \de^2)/2 - (\al\de - \bt\ga) \nonumber \\
&= &(\al^2 + 2\bt\ga + \de^2)/2. \label{a6} \ea
Formula (\ref{a6}) is referenced in the discussion of the Schmidt decomposition (App. B).

By solving the equations $ME_\pm = \la_\pm E_\pm$ and $M^tF_\pm = \la_\pm F_\pm$, one finds that the eigenvectors
\be E_\pm \propto \left[\begin{array}{c} 1 \\ (\la_\pm - \al)/\bt \end{array}\right]
\propto \left[\begin{array}{c} (\la_\pm - \de)/\ga \\ 1 \end{array}\right] \label{a7} \ee
and the adjoint eigenvectors
\be F_\pm \propto \left[\begin{array}{c} 1 \\ (\la_\pm - \al)/\ga \end{array}\right]
\propto \left[\begin{array}{c} (\la_\pm - \de)/\bt \\ 1 \end{array}\right]. \label{a8} \ee
For definiteness, we choose
\be E_\pm = N_\pm \left[\begin{array}{c} 1 \\ e_\pm \end{array}\right], \ \ 
F_\pm = \pm N_\pm \left[\begin{array}{c} f_\pm \\ 1 \end{array}\right], \label{a9} \ee
where $e_\pm = (\la_\pm - \al)/\bt$, $f_\pm = (\la_\pm - \de)/\bt$ and $N_\pm$ are normalization constants.
It is easy to verify that $E_+^tF_+ \propto e_+ + f_+ = 2\si/\bt$, where $\si = [(\al + \de)^2/4 - (\al\de - \bt\ga)]^{1/2}$. Likewise, $E_-^tF_- \propto -(e_-  + f_-) = 2\si/\bt$. Hence, $N_\pm = N = (\bt/2\si)^{1/2}$. It is also easy to verify that $F_\pm^tE_\mp \propto f_\pm +  e_\mp = (\la_\pm - \de)/\bt + (\la_\mp - \al)/\bt = 0$, because $\la_+ + \la_- = \al + \de$. Hence, the direct and adjoint eigenvectors are orthonormal.

Let $E$ be the $2 \times 2$ matrix whose columns are $E_+$ and $E_-$, and $F$ be the matrix whose columns are $F_+$ and $F_-$. Then the orthonormality equation is $E^tF = I = F^tE$. Written explicitly, the product matrix
\ba E^tF &= &N^2 \left[\begin{array}{cc} 1 & e_+ \\ 1 & e_- \end{array}\right]
\left[\begin{array}{cc} f_+ & -f_- \\ 1 & -1 \end{array}\right] \nonumber \\
&= &{\bt \over 2\si} \left[\begin{array}{cc} e_+ + f_+ & -e_+ - f_- \\ e_- + f_+ & -e_- - f_- \end{array}\right]. \label{a10} \ea
The related product
\ba EF^t &= &N^2 \left[\begin{array}{cc} 1 & 1 \\ e_+ & e_- \end{array}\right]
\left[\begin{array}{cc} f_+ & 1 \\ -f_- & -1 \end{array}\right] \nonumber \\
&= &{\bt \over 2\si} \left[\begin{array}{cc} f_+ - f_- & 0 \\ e_+f_+ - e_-f_- & e_+ - e_- \end{array}\right]. \label{a11} \ea
It is easy to verify that $e_+ - e_- = f_+ - f_- = 2\si/\bt$ and $e_+f_+ = e_-f_- = \ga/\bt$. Hence, $EF^t = I$, as required by the completeness equation. Although the matrices $E^tF$ and $EF^t$ are formally distinct, they both equal $I$.
Now let $\La = \diag(\la_+,\la_-)$. Then
\ba E\La F^t &= &N^2 \left[\begin{array}{cc} 1 & 1 \\ e_+ & e_- \end{array}\right]
\left[\begin{array}{cc} \la_+ & 0 \\ 0 & \la_- \end{array}\right]
\left[\begin{array}{cc} f_+ & 1 \\ -f_- & -1 \end{array}\right] \nonumber \\
&= &{\bt \over 2\si} \left[\begin{array}{cc} \la_+f_+ - \la_-f_- & \la_+ - \la_- \\ e_+\la_+f_+ - e_-\la_-f_- & e_+\la_+ - e_-\la_- \end{array}\right]. \label{a12} \ea
By using the preceding identities, it is easy to verify that $\la_+ - \la_- = 2\si$, $e_+\la_+ - e_-\la_- = 2\si\de/\bt$, $f_+\la_+ - f_-\la_- = 2\si\al/\bt$ and $e_+\la_+f_+ - e_-\la_-f_- = (\ga/\bt)(\la_+ - \la_-) = 2\si\ga/\bt$. Hence, $E\La F^t = M$, as required by the decomposition theorem.

The second example concerns the generating matrix
\be G = \left[\begin{array}{cc} k_1 & k_2 - k_3 \\ k_2 + k_3 & -k_1 \end{array}\right], \label{a13} \ee
which has the eigenvalues $\la_\pm = \pm k$, where $k = (k_1^2 + k_2^2 - k_3^2)^{1/2}$. The direct and adjoint eigenvectors can be written in the forms of Eqs. (\ref{a9}), where $e_\pm = (\pm k - k_1)/(k_2 - k_3)$, $f_\pm = (\pm k + k_1)/(k_2 - k_3)$ and $N = [(k_2 - k_3)/2k]^{1/2}$. The adjoint decomposition of the generating matrix is important, because if $G = E\La_gF^t$, then $G^n = E\La_g^nF^t$. Hence, the exponential $M = e^G = E\La_mF^t$, where $\La_m = \diag(\la_m)$ and $\la_m = e^{\la_g}$.

It follows from the symplectic generator equation (\ref{3.4}) that if $\la$ is an eigenvalue of $G$, with eigenvector $E$, then $-\la$ is an eigenvalue of $G^t$, with eigenvector $JE$. But $G$ and $G^t$ have the same eigenvalues, so the eigenvalues of $G$ occur in positive and negative (or zero) pairs. Hence, $\tr(G) = 0$ and $\det(M) = 1$. If $\det(G) \neq 0$ (as in the second example), then the eigenvalues are all nonzero.

\newpage

\sec{Appendix B: Schmidt decomposition}

In this appendix, we discuss the Schmidt decomposition of a symplectic matrix.
Every real matrix $M$ has the Schmidt decomposition $QDP^t$, where $D$ is diagonal and nonnegative, and $P$ and $Q$ are orthogonal \cite{hor13}. The columns of $P$ (input vectors) are the eigenvectors of $M^tM$, the columns of $Q$ (output vectors) are the eigenvectors of $MM^t$ and the entries of $D$ (Schmidt coefficients) are the square roots of the (common) eigenvalues of $M^tM$ and $MM^t$. If $M$ is symmetric, then $P = Q$ and the entries of $D$ are the eigenvalues of $M$.

For the symplectic matrix (\ref{a3}), the product matrices
\ba M^tM &= &\left[\begin{array}{cc} \al^2 + \ga^2 & \al\bt + \ga\de \\ \al\bt + \ga\de & \bt^2 + \de^2 \end{array}\right], \label{b5} \\
MM^t &= &\left[\begin{array}{cc} \al^2 + \bt^2 & \al\ga + \bt\de \\ \al\ga + \bt\de & \ga^2 + \de^2 \end{array}\right]. \label{b6} \ea
Both matrices are symmetric and have determinant 1 (as they should do). It follows from Eq. (\ref{a4}) that they share the eigenvalues (squared Schmidt coefficients)
\be \si_\pm^2 = A \pm (A^2 - 1)^{1/2}, \label{b7} \ee
where $A = (\al^2 + \bt^2 + \ga^2 + \de^2)/2$. It is easy to verify that $\si_+^2\si_-^2 = 1$, so the Schmidt coefficients represent stretching ($+$) and squeezing ($-$). By comparing Eqs. (\ref{a5}), (\ref{a6}) and (\ref{b7}), one finds that the coefficients are similar to, but different from, the eigenvalues ($\bt^2 + \ga^2 \neq 2\bt\ga$). By writing $M$ in the form of Eq. (\ref{2.31}), one finds that $A = C^2 + S^2 = \cosh(2\la)$ and $\si_\pm^2 = \cosh(2\la) \pm \sinh(2\la) = e^{\pm 2\la}$. Hence, the Schmidt coefficients are the stretching and squeezing factors of the dilation. Area in state space is conserved because these factors are reciprocal.

It follows from Eq. (\ref{b5}) that the first product matrix
\be M^tM = \left[\begin{array}{cc} A' + \De' & B' \\ B' & A' - \De' \end{array}\right]
= \left[\begin{array}{cc} C + Sc & Ss \\ Ss & C - Sc \end{array}\right], \label{b8} \ee
where $A' = A$, $B' = \al\bt + \ga\de$ and $\De' = (\al^2 + \ga^2 - \bt^2 - \de^2)/2$.
The second form of Eq. (\ref{b8}) follows from Eqs. (\ref{2.6}) and (\ref{2.14}). In this representation, $C = A'$, $S = [(B')^2 + (\De')^2]^{1/2} = (C^2 - 1)^{1/2}$ and the dilation angle
\be \tan(2\th_1) = B'/\De' = 2(\al\bt + \ga\de)/(\al^2 - \bt^2 + \ga^2 - \de^2). \label{b9} \ee
It follows from Eq. (\ref{b6}) that the second product matrix also can be written in the forms of Eq. (\ref{b8}), with $B' = \al\ga + \bt\de$ and $\De' = (\al^2 + \bt^2 - \ga^2 - \de^2)/2$. Hence, the dilation angle
\be \tan(2\th_2) = B'/\De' = 2(\al\ga + \bt\de)/(\al^2 + \bt^2 - \ga^2 - \de^2). \label{b10} \ee
Equations (\ref{b9}) and (\ref{b10}) are identical to Eqs. (\ref{2.41}) and (\ref{2.42}), respectively. They are also equivalent to Eqs. (\ref{3.55}).

To summarize these results, the product matrices have the same Schmidt coefficients (as they should do), but different dilation directions. The output vectors are rotated versions of the input vectors: $P = R(\th_1)$ and $Q = R(\th_2) = R(\th_2 - \th_1)R(\th_1)$, where $R$ is a rotation matrix. The input vectors (columns of $P$) are orthogonal to each other, as are the output vectors (columns of $Q$). However, $QP^t = R(\th_2 - \th_1)$, so the input and output vectors are orthogonal if and only if $\th_1 = \th_2$. Notice that the components of the vectors are sines and cosines of the angles $\th_1$ and $\th_2$, whereas Eqs. (\ref{b9}) and (\ref{b10}) involve the double angles $2\th_1$ and $2\th_2$.

In the preceding text, we derived formulas for the Schmidt decomposition of a $2 \times 2$ symplectic matrix. Some results for $2n \times 2n$ symplectic matrices, which follow from Eqs. (\ref{2.1}) and (\ref{2.2}), are worth mentioning \cite{mck13}. If $M$ is symplectic, then so are $M^t$ and $M^{-1}$. If $M_1$ and $M_2$ are symplectic, then so is their product $M_2M_1$. In particular, the product matrices $M^tM$ and $MM^t$ are symplectic (and symmetric). If $E$ is an eigenvector of $M$ with (nonzero) eigenvalue $\la$, then $JE$ is an eigenvector of $M^t$, with eigenvalue $1/\la$ (because $M^{-1} = J^tM^tJ$ and $JJ^t = I$). $M$ and $M^t$ have the same eigenvalues, so $1/\la$ is also an eigenvalue of $M$. Hence, the eigenvalues of a symplectic matrix occur in reciprocal pairs. If $E$ is an (input) eigenvector of $M^tM$ with (nonzero) eigenvalue $\si^2$, then $JE$ is an eigenvector of $(M^tM)^t = M^tM$ with eigenvalue $1/\si^2$. A similar statement can be made about the (output) eigenvectors and eigenvalues of $MM^t$. Furthermore, if $E$ is an eigenvector of $M^tM$, then $F = ME$ is an eigenvector of $MM^t$, with the same eigenvalue. Conversely, if $F$ is a eigenvector of $MM^t$, then $E = M^tF$ is an eigenvector of $M^tM$. Hence, the product matrices have the same eigenvectors (squared Schmidt coefficients), which occur in reciprocal pairs.

\newpage

\sec{Appendix C: Asymmetric matrix products}

In Sec. 4, we considered the product of two symmetric (dilation) matrices. In this appendix, we consider the product of two asymmetric (dilation--rotation) matrices.
Each matrix has the Schmidt decomposition $M_i = Q_iD_iP_i^t$, where $P_i = R(\ph_i)$ and $Q_i = R(\th_i)$. The composite transformation is
\be Y = (Q_2D_2P_2^t)(Q_1D_1P_1^t)X. \label{c1} \ee
The first matrix $P_1^t$ represents a rotation of the input axes, whereas the last matrix $Q_2$ represents a rotation of the output axes. In terms of rotated axes, the transformation is
\be (Q_2^tY) = D_2(P_2^tQ_1)D_1(P_1^tX). \label{c2} \ee
The intermediate matrix $M_3 = D_2(P_2^tQ_1)D_1$ has the decomposition $Q_3D_3P_3^t$, where the dilation parameter $\la_3$, and the angles $\ph_3$ and $\th_3$, depend on the constituent dilation parameters $\la_1$ and $\la_2$, and the difference angle $\ps = \ph_2 - \th_1$. In terms of the original axes,
\be Y = (Q_2Q_3)D_3(P_1P_3)^tX. \label{c3} \ee
The composite matrix $M_4 = (Q_2Q_3)D_3(P_1P_3)^t$ has the decomposition $Q_4D_4P_4^t$, where the composite dilation parameter $\la_4 = \la_3$, and the composite angles $\ph_4 = \ph_1 + \ph_3$ and $\th_4 = \th_2 + \th_3$.
Notice that the way in which the transformations combine is determined by the intermediate matrix, which is relatively simple.

Written explicitly, the intermediate matrix
\ba M_3 &= &\left[\begin{array}{cc} E_2 & 0 \\ 0 & 1/E_2 \end{array}\right]
\left[\begin{array}{cc} c & s \\ -s & c \end{array}\right]
\left[\begin{array}{cc} E_1 & 0 \\ 0 & 1/E_1 \end{array}\right] \nonumber \\
&= &\left[\begin{array}{cc} E_2 & 0 \\ 0 & 1/E_2 \end{array}\right]
\left[\begin{array}{cc} cE_1 & s/E_1 \\ -sE_1 & c/E_1 \end{array}\right] \nonumber \\
&= &\left[\begin{array}{cc} cE_2E_1 & sE_2/E_1 \\ -sE_1/E_2 & c/E_2E_1 \end{array}\right], \label{c4} \ea
where $E_i = \exp(\la_i)$, $c = \cos(\ps)$ and $s = \sin(\ps)$. Let $\la_\pm = \la_2 \pm \la_1$, $C_\pm = \cosh(\la_\pm)$ and $S_\pm = \sinh(\la_\pm)$. Then
\be M_3 = \left[\begin{array}{cc} c(C_+ + S_+) & s(C_- + S_-) \\ -s(C_- - S_-) & c(C_+ - S_+) \end{array}\right]. \label{c5} \ee
Notice that matrix (\ref{c5}) is symmetric (and represents a dilation) if and only if $sC_- = 0$, which requires that $\ps = 0$ or $\pi$ (parallel or antiparallel dilations). For other difference angles, the matrix is asymmetric (and represents a dilation followed by a rotation).
By comparing Eq. (\ref{c5}) to Eq. (\ref{2.31}), one finds that
\be C_3c_- = cC_+, \ \ S_3s_+ = sS_-, \ \ C_3s_- = -sC_-, \ \ S_3c_+ = cS_+, \label{c6} \ee
where $C_3 = \cosh(\la_3)$ and $c_\pm = \cos(\th_3 \pm \ph_3)$. The definitions of $S$ and $s_\pm$ are similar.

It follows from Eqs. (\ref{c6}) that
\ba C_3^2 &= &c^2C_+^2 + s^2C_-^2 \nonumber \\
&= &c^2(C_2C_1 + S_2S_1)^2 + s^2(C_2C_1 - S_2S_1)^2 \nonumber \\
&= &C_2^2C_1^2 + S_2^2S_1^2 + 2C_2C_1S_2S_1(c^2 - s^2), \label{c7} \\
S_3^2 &= &c^2S_+^2 + s^2S_-^2 \nonumber \\
&= &c^2(S_2C_1 + C_2S_1)^2 + s^2(S_2C_1 - C_2S_1)^2 \nonumber \\
&= &S_2^2C_1^2 + C_2^2S_1^2 + 2C_2C_1S_2S_1(c^2 - s^2). \label{c8} \ea
Equations (\ref{c7}) and (\ref{c8}) are consistent with Eqs. (\ref{4.10}) and (\ref{4.11}). The only difference between the equations is the replacement of $\th_2 - \th_1$ by $\ph_2 - \th_1$. Hence, the results described by Fig. 1 are~general.
With $C_3$ and $S_3 \ge 0$ known, the dilation parameter $\la_3 = \log(C_3 + S_3)$.
Alternatively, by using the double-argument formulas for hyperbolic trigonometric functions, one can rewrite Eqs. (\ref{c7}) and (\ref{c8}) in the compact form
\be D_3 = c^2D_+ + s^2D_- = D_2D_1 + T_2T_1d, \label{c9} \ee
where $D_i = \cosh(2\la_i)$, $T_i = \sinh(2\la_i)$ and $d = \cos(2\ps)$.

It also follows from Eqs. (\ref{c6}) that
\be \tan(\th_3 + \ph_3) = rS_-/S_+, \ \ \tan(\th_3 - \phi_3) = -rC_-/C_+, \label{c10} \ee
where $r = s/c$. Notice that $d$ and $r$ have the common domain $[-\pi/2, \pi/2)$. The input and output angles
\ba \tan(2\ph_3) &= &{2rT_2 \over T_+ - r^2T_-}
\ = \ {tT_2 \over D_2T_1 + T_2D_1d}, \label{c11} \\
\tan(2\th_3) &= &{-2rT_1 \over T_+ + r^2T_-}
\ = \ {-tT_1 \over T_2D_1 + D_2T_1d}, \label{c12} \ea
where $t = \sin(2\ps)$.
One obtains the second parts of Eqs. (\ref{c11}) and (\ref{c12}) from their first parts by multiplying their numerators and denominators by $c^2$, and using double-angle formulas.
If the dilations are aligned ($r = 0$), then $\ph_3 = \th_3 = 0$. If the dilations have equal magnitudes ($\la_- = 0$), then $\tan(2\ph_3) = r/\cosh(2\la)$ and $\tan(2\th_3) = -r/\cosh(2\la)$. With the intermediate angles known, the composite angles $\ph_4 = \ph_1 + \ph_3$ and $\th_4 = \th_2 + \th_3$, as stated above.

For the special case in which $M_1$ and $M_2$ are both dilations ($\th_i = \ph_i$ and $\ps = \th_{21}$), it is convenient to work in the rotated frame [Eq. (\ref{4.7})]. By comparing the composite matrix $M_4 = R_{21}D_2R_{21}^tD_1$ to the intermediate matrix $M_3 = D_2R_{21}^tD_1$, one finds that $\la_4 = \la_3$, $\ph_4 = \ph_3$ and $\th_4 = \th_{21} + \th_3$. It follows from Eqs. (\ref{c10}) that the composite sum and difference angles
\ba \tan(\th_4 + \ph_4) &= &{r(S_+ + S_-) \over S_+ - r^2S_-}
\ = \ {S_2C_1t \over C_2S_1 + S_2C_1d}, \ \ \label{c21} \\
\tan(\th_4 - \ph_4) &= &{r(C_+ - C_-) \over C_+ + r^2C_-} 
\ = \ {S_2S_1t \over C_2C_1 + S_2S_1d}.\ \ \label{c22} \ea
Equations (\ref{c21}) and (\ref{c22}) are equivalent to Eqs. (\ref{4.12}). (In Sec. 4, we used the notation $d = c_{21}$ and $t = s_{21}$.) It is straightforward to verify that Eqs. (\ref{c11}) and (\ref{c12}) are consistent with the first of Eqs. (\ref{4.13}) and Eq. (\ref{4.14}), respectively. (In Sec. 4, we denoted the composite output angle $\th_4$ by $\th_3$, because there was no intermediate matrix.)

\end{document}